\documentstyle[10pt,fps]{article}

\begin{document}

\centerline{\large \bf Spectroscopy and large scale wave functions
}

\vskip 1.cm
\centerline{Qi-Zhou Chen,$^{\rm{a}}$ Shuo-Hong Guo, $^{\rm{a}}$,
  Xiang-Qian Luo, $^{\rm{b}}$ and Antonio J. Segui-Santonja $^{\rm{c}}$}

\vskip 1.cm

  \centerline{\it $^{\rm{a}}$ CCAST
    (World Laboratory),  P.O. Box 8730, Beijing 100080, China}

  \centerline{\it  and Department of Physics, Zhongshan
                                        University, Guangzhou 510275, China
                                        }
\centerline{\it $^{\rm{b}}$ HLRZ, Forschungszentrum,
D-52425 J{\"u}lich, Germany}
\centerline{\it and Deutsches Elektronen-Synchrotron DESY, D-22603 Hamburg,
Germany}

\centerline{\it $^{\rm{c}}$ Departamento de F{\'\i}sica Te{\'o}rica,
Universidad de Zaragoza, 50009 Zaragoza, Spain
}

\vskip 1.cm

\begin{abstract}
  We discuss the relevance of long wavelength excitations for the low energy
  spectrum of \rm{QCD}, and try to develop an efficient method for solving
  the Schr{\"o}dinger equation, and for extracting the glueball masses and long
  wavelength functions of the ground and excited states.
  Some technical problems appearing in the calculations of
  \rm{SU(3)} gauge theory are
  discussed.
\end{abstract}

\vskip 1.cm

QCD in the pure gauge sector possesses a nontrivial vacuum structure
and bound states called glueballs.
Solving the
Schr{\"o}dinger equation in the Hamiltonian formulation
can directly provide not only the glueball masses from the eigenvalues,
but also the profiles, i.e.,
the wave functions for the ground state and excited states.

Strictly speaking,
the continuum physics should be extracted in the
asymptotic scaling region predicted by the renormalization group equation.
In \cite{GCL,CGZF,QCD3,MASS,ASSYM}, we developed
an efficient eigenvalue equation method with
some new truncation schemes which preserve the continuum limit.
This is an essential step towards the scaling.

Our starting point is to obtain the long wavelength wave functions
of the ground and excited states.
The philosophy is to keep the correct long
wavelength limit during the truncation.
The low energy spectrum originates mainly from the long wavelength excitations.
This is because
the size or the Compton length of a glueball, which
is usually of the same order as that of a hadron
or the lattice size, should be much greater than the lattice spacing $a$.
It is worth mentioning that the confinement of gluons or static quarks
is closely related to the long wavelength structure of the vacuum.
For the long wavelength
configurations $U$, the ground state in the continuum \cite{Arisue} is
\begin{eqnarray*}
\vert \Omega \rangle=exp \lbrack - {\mu_0 \over e^2} \int d^{D-1}x ~ tr {\cal
F}^2
\end{eqnarray*}
\begin{eqnarray}
- {\mu_2 \over e^6} \int d^{D-1}x  ~tr ({\cal D} {\cal F})^2 \rbrack,
\label{a1}
\end{eqnarray}
with $e$ being the renormalized coupling,
${\cal F}$ the field strength tensor and ${\cal D}$ the covariant derivative.

Our method is as follows.
The ground state
\begin{eqnarray}
  \vert \Omega \rangle = exp \lbrack R(U) \rbrack \vert 0 \rangle
\label{b1}
\end{eqnarray}
with
energy $\epsilon_{\Omega}$ of the Hamiltonian $H$
has to satisfy the Schr{\"o}dinger
equation
$H \vert \Omega \rangle = \epsilon_{\Omega} \vert \Omega \rangle$,
which results in
\begin{eqnarray*}
\sum_{l} \lbrace [E_l,[E_l,R(U)]]+[E_l,R(U)][E_l,R(U)] \rbrace
\end{eqnarray*}
\begin{eqnarray}
- {2 \over g^4} \sum_{p} tr(U_p+U_{p}^{\dagger})
={2a \over g^2} \epsilon_{\Omega}.
\end{eqnarray}
This equation can be solved by a systematic method
\cite{GCL,CGZF,QCD3,MASS,ASSYM},
in which $R(U)$ is expanded in order of graphs $G_{n,i}$, i.e.,
\begin{eqnarray}
  R(U)=\sum_{n} R_{n}(U)=\sum_{n,i} C_{n,i} G_{n,i}(U),
\end{eqnarray}
with
$n$ being the order of the graphs defined as the number of plaquettes involved.
Then the $N$th order truncated eigenvalue equation is
\begin{eqnarray*}
\sum_{l} \lbrace [E_l,[E_l,\sum_{n}^{N} R_{n}(U)]]
\end{eqnarray*}
\begin{eqnarray*}
+\sum_{n_1+n_2 \le N}[E_l,R_{n_1}(U)][E_l,R_{n_2}(U)] \rbrace
\end{eqnarray*}
\begin{eqnarray}
- {2 \over g^4} \sum_{p} tr(U_p+U_{p}^{\dagger})
={2a \over g^2} \epsilon_{\Omega},
\label{b2}
\end{eqnarray}
from which the
coefficients $C_{n,i}$ are determined. Similar equation for the glueball
mass and its wave function can be derived \cite{MASS}.

Only the second term generates new or higher order graphs
(of order $n_1+n_2$),
influencing the choice of independent graphs.
The essential feature of our truncation scheme,
which differs sufficiently from the scheme in \cite{Green},
is in the treatment of this second term.
It has been generally proven \cite{GCL} that in the long wavelength limit
this term should behave as
\begin{eqnarray}
  [E_l,R_{n_1}(U)][E_l,R_{n_2}(U)] \propto a^6 ~tr({\cal D} {\cal
F}_{\mu,\nu})^2.
\end{eqnarray}
To preserve this correct limit, when the equation
(\ref{b2}) is truncated to the $Nth$ order all the graphs created by
$[E_l,R_{n_1}(U)][E_l,R_{n_2}(U)]$ for $n_1+n_2 \le N$ must be considered.
In this method, no group integration is necessary, and
independent graphs can be obtained systematically \cite{QCD3,MASS}, which can
be
suitably chosen for more rapid convergence to the continuum limit.

When the eigenvalue equation is truncated at a finite order, some ambiguity
appears, as mentioned in \cite{QCD3,GCFC}.
For a non-abelian gauge theory, the unimodular
condition induces the mixing of a graph with not only the same order, but also
different orders. Therefore, the definition of the order of a graph and the
choice of independent graphs are not
unique. In the infinite order limit, different prescriptions should give
identical results. At low orders, this is not necessarily the case.
Then a question arises: which classification scheme
works more efficiently and converges
more rapidly to the continuum limit?
(In numerical simulation on a finite lattice,
there is analogously
a problem of choosing better operators or wave functions).

In SU(2) gauge group,
$TrU_p=TrU_{p}^{\dagger}$ and all loops with crossing can
be transformed into loops without crossing.
The complication of the $\rm{SU(3)}$ group, however,
is that not all the disconnected graphs can be
transformed
to the connected ones.
 Any group element $A$ of SU(3)
has to satisfy the following condition
\begin{eqnarray}
\label{h1}
A_{i_1j_1}A_{i_2j_2}A_{i_3j_3}\epsilon_{j_1j_2j_3}=\epsilon_{i_1i_2i_3},
\end{eqnarray}
where a summation over the repeated indices is implied. We rewrite
this condition as
\begin{eqnarray*}
  2(A^{\dagger})_{ij}=2(A^2)_{ij}-2A_{ij}trA
\end{eqnarray*}
\begin{eqnarray}
  +[(trA)^2-tr(A^2)]\delta_{ij},
\end{eqnarray}
or
\begin{eqnarray*}
  2\delta_{ij}=2(A^3)_{ij}-2(A^2)_{ij}trA
\end{eqnarray*}
\begin{eqnarray}
  + [(trA)^2-tr(A^2)]A_{ij},
\end{eqnarray}
from which the relations among different graphs can be established.
One sees that not only graphs of the same order, but also
graphs of different orders mix,  so that the classification becomes
rather involved.

In \cite{GCL,CGZF}, the disconnected graphs were taken as independent graphs.
We realize that this is not the most efficient way, because the disconnected
graphs have large overlapping with graphs of lower orders in the weak
coupling region. Indeed,
the results in \cite{CGZF} showed that such a choice was not so economical.
Alternatively, the connected graphs represent more coherence and have less
mixing with lower order graphs.
The superiority of the connected scheme over
the disconnected one has also been shown in \cite{GCFC} for a
(2+1)-dimensional $\rm{SU(2)}$ model.
This is why for a more realistic gauge group $\rm{SU(3)}$,
we try our best to transform \cite{QCD3} the disconnected
graphs to the connected ones.

\begin{figure}[htb]
\fpsxsize=7.5cm
\vspace{-25mm}
\fpsbox[70 90 579 760]{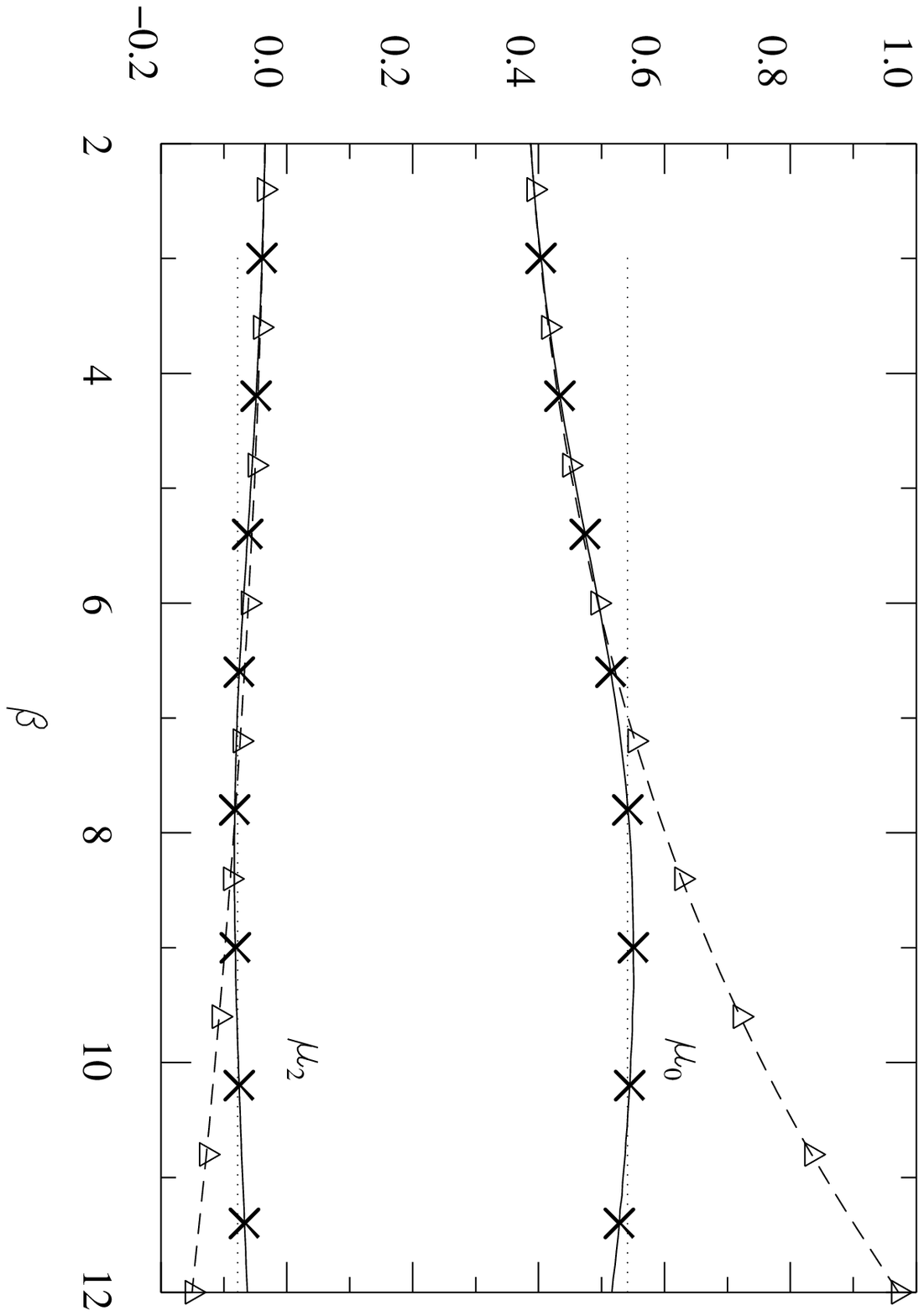}
\vspace{-10mm}
\caption{Coefficients in the vacuum wave function of $\rm{QCD}_3$.
Triangles: strong coupling
expansion; Crosses: the third order
calculation described in the text;
The dot lines: mean values in the scaling
region.}
\label{fig1}
\end{figure}

The coefficients
$\mu_0$ and $\mu_2$ in (\ref{a1}) should be constants in the weak coupling
limit
$g \to 0$ or $\beta=6/g^2 \to \infty$ as required by the renormalizability
of the theory.
Figure \ref{fig1} shows the third order results for $\rm{QCD_3}$ from the
strong coupling expansion and the truncated eigenvalue equation.
They are consistent in the strong coupling region.
For larger $\beta$, it is not surprising that
the strong coupling expansion method
no longer works. It is usually hoped that beyond the strong coupling
region, there is
a scaling region for extracting continuum information when
the physical quantities become approximately constants.
{}From the intermediate coupling till
the weak coupling, the data from the
the truncated eigenvalue equation method show nice scaling behavior,
thus suggesting the correct long wavelength continuum limit (\ref{a1})
of the vacuum wave
function (\ref{b1}).

In \cite{ASSYM}, we present a possibly more efficient, named ``inverse
scheme''.
Define
the linear operations of derivation $Dev$ and inverse $Inv$ as
\begin{eqnarray*}
  Dev[G]=\sum_{l} [E_l,[E_l,G]],
\end{eqnarray*}
\begin{eqnarray}
  Inv[Dev[G]]=G.
  \label{DEF}
\end{eqnarray}
Let us take some of the graphs in \cite{MASS,QCD3} as examples,
\begin{eqnarray*}
Dev[G_{1}^{\dagger}]={16 \over 3} G_{1}^{\dagger},
\end{eqnarray*}
\begin{eqnarray*}
Inv[G_{1}^{\dagger}]= {3 \over 16} G_{1}^{\dagger},
\end{eqnarray*}
\begin{eqnarray}
  Dev[G_{2,1}]={40 \over 3} G_{2,1} + 8 G_{1}^{\dagger}.
  \label{EXAM}
\end{eqnarray}
Using (\ref{DEF}) and (\ref{EXAM}),
we have
\begin{eqnarray}
  Inv[G_{2,1}]={3 \over 40}( G_{2,1} - {3 \over 2} G_{1}^{\dagger}).
\end{eqnarray}

In the inverse scheme,
we choose the inverse of graphs $G_{n,j}$

\begin{eqnarray}
  G_{n,i}^{I} \propto Inv[G_{n,j}]
\end{eqnarray}
as
independent graphs of order $n$.
In general, $G_{n,i}^{I}$ is a linear combination of $G_{n,j}$
and lower order graphs.  Up to the third order, the inverse of
connected overlapping  (c.o.) graphs $G_{n,j}(c.o.)$ in \cite{QCD3,MASS}
 are

\begin{eqnarray*}
G_{2,1}^{I} =  G_{2,1}-{3 \over 2} G_{1}^{\dagger},
\end{eqnarray*}
\begin{eqnarray*}
G_{3,1}^{I} =  G_{3,1}-G_{2,2},
\end{eqnarray*}
\begin{eqnarray*}
G_{3,3}^{I}= G_{3,3}- {9 \over 28} G_{2,5}+ {9 \over 28} G_{2,6},
\end{eqnarray*}
\begin{eqnarray}
G_{3,6}^{I}=G_{3,6}- {1 \over 18} G_{3,5} - {41 \over 99} G_{2,4} + {227 \over
396} G_{2,3}.
\label{sub}
\end{eqnarray}

We observe that in the inverse scheme, the independent operators $G_{n,i}^{I}$
are given by $G_{n,j}$ with subtraction of lower order graphs.
(Actually there is alternation in sign in the lower order
graphs in (\ref{sub}), but the dominant ones are subtractions).
Such a subtraction may further reduce the overlapping of higher order graphs
with lower order ones in the weak coupling region. Therefore, we
expect that the inverse scheme may be superior
over the other schemes.

Whereas our goal is to apply this method to QCD in 4 dimensions (in progress),
we would like to demonstrate how the inverse scheme works in
an  interesting and relevant but less complicated theory: $\rm{QCD}_3$,
the $3D$ $\rm{SU(3)}$ gauge model.
This reduces a theory to a super-renormalizable one, since the
renormalization requirements amounts to dimensional analysis.
In the weak coupling region (for large $\beta=6/g^2$),
because the renormalized charge $e$ and the bare coupling are related by
$g^2=e^2 a$, dimensional analysis tells us that the dimensionless masses
$aM_{J^{PC}}$
should scale as
\begin{eqnarray}
{aM_{J^{PC}} \over g^2} \to {M_{J^{PC}} \over e^2} \approx const.,
\end{eqnarray}
from which the physical glueball masses $M_{J^{PC}}$ are obtained.

\begin{figure}[htb]
\fpsxsize=7.5cm
\vspace{-25mm}
\fpsbox[70 90 579 760]{mass_qcd3.ps}
\vspace{-10mm}
\caption{Glueball masses in $\rm{QCD}_3$ indicated by crosses.
The dot lines: mean values in the scaling
region; Dash lines: strong coupling expansion;
Triangles: results from a connected scheme.}
\label{fig2}
\end{figure}

The third order results for
$aM_{0^{++}}/g^2$ and $aM_{0^{--}}/g^2$ are shown in Fig. \ref{fig2} by
crosses.
For comparison, the results \cite{MASS} using the old classification scheme
(triangles) truncated up to the same order
and those from the strong coupling expansion (dash lines)
are also included in the figure.
One sees clearly the advantage of the inverse scheme.
While the mean values from different schemes are consistent,
the scaling window for $aM_{0^{++}}/g^2$ from the old scheme is $\beta \in
[5,8)$, and that
from the inverse scheme is greatly widen and extended to
much weaker coupling $\beta \approx 12$.
The mean values for the masses extracted in the scaling region are

\begin{eqnarray*}
  {M_{0^{++}} \over e^2} \approx 2.0923 \pm 0.0334,
\end{eqnarray*}
\begin{eqnarray}
  {M_{0^{--}} \over e^2} \approx 3.7077 \pm 0.0280.
\end{eqnarray}

In this paper, we discuss only the results for $\rm{QCD_3}$.
The results of $3D-\rm{U(1)}$,
and $\rm{SU(2)}$ gauge theories and $2D-\sigma$ model
have been presented in
\cite{GCL,CGZF,GCFC} and summarized in \cite{Guo}.
Even at low orders,
clear scaling
windows
for the physical quantities in all cases
have been established, where the results
are in perfect
agreement with the Monte Carlo data \cite{Arisue,Teper}
(they have data only for $3D$ $SU(2)$).
Extension to 3+1 dimensional nonabelian guage theories is in progress.

To ensure that the results occur at their correct values,
efforts have to be made in doing higher order calculations.
In the applications to $3D-\rm{U(1)}$ and $2D-\sigma$ models \cite{Liu,Guo},
the convergence has been confirmed.
In conclusion, the method
is hopeful to be developed into an efficient systematic approach to
extracting the continuum physics.

QZC and SHG are supported by Institute of Higher Education, and
XQL is sponsored by DESY. We would like to thank
H. Arisue, P.F. Cai, X.Y. Fang, C. Hamer,
L. Li, J.M. Liu, D. Sch{\"u}tte, and W.H. Zheng for useful discussions.

\end{document}